\documentclass[a4paper, fleqn, usenatbib, onecolumn]{mnras}

\usepackage{newtxtext, newtxmath}

\usepackage[T1]{fontenc}
\usepackage{ae, aecompl}
\usepackage{color}
\def\comment#1{}

\usepackage{graphicx}	
\usepackage{amsmath}	
\usepackage{amssymb}	

\title[Investigating eccentricities of the binary black hole
    .  .  .  ]{Investigating eccentricities of the binary black hole signals from the LIGO-Virgo catalog GWTC-1}

\author[Q-Y Yun, W-B Han, G Wang, S-C Yang]{
Qian-Yun Yun$^{1,2}$,  Wen-Biao Han$^{1,3}$\thanks{E-mail: wbhan@shao.ac.cn}, Gang Wang$^{1,3}$\thanks{E-mail: gwang@shao.ac.cn}, Shu-Cheng Yang$^{1,3}$
\\
1. Shanghai Astronomical Observatory,  Chinese Academy of Sciences,  Shanghai,  China,  200030 \\
2. School of Physical Science and Technology, ShanghaiTech University, Shanghai,  China, 201210 \\
3. School of Astronomy and Space Science,  University of Chinese Academy of Sciences,  Beijing,  100049,  China}

\pubyear{2020}

\begin{document}
\label{firstpage}
\pagerange{\pageref{firstpage}--\pageref{lastpage}}
\maketitle

\begin{abstract}
In the first Gravitational-Wave Transient Catalogue of LIGO and Virgo,  all events are announced having zero eccentricity.   In the present paper,  we investigate the performance of SEOBNRE which is a spin-aligned eccentric waveform model in time-domain.   By comparing with all the eccentric waveforms in SXS library,  we find that the SEOBNRE coincides perfectly with numerical relativity data.   Employing the SEOBNRE,  we re-estimate the eccentricities of all black hole merger events. We find that most of these events allow a possibility for existence of initial eccentricities at 10 Hz band,  but are totally circularized at the observed frequency ($ \gtrsim 20$ Hz).   The upcoming update of LIGO and the next generation detector like as Einstein Telescope,  will observe the gravitational waves starting at 10 Hz or even lower.   If the eccentricity exists at the lower frequency,  it may  significantly support the dynamical formation mechanism taking place in globular clusters.  
\end{abstract}



\section{introduction}

The successful detection of gravitational waves (GWs) by Advanced LIGO and Virgo (\cite{abbott2016tests,abbott2016binary,abbott2016characterization,abbott2016gw151226,abbott2016astrophysical,abbott2017exploring,abbott2017gw170608,abbott2019gwtc,abbott2019tests,abbott2019binary}) announces that the era of GW Astronomy is coming.   Along with the more and more events will be found in the future,  the formation mechanics and distribution of GW events might be revealed.   
Up to now,  The Advanced LIGO and Virgo has announced 12 gravitational wave events,  most of them (10) are coalescence of binary black holes (\cite{abbott2019gwtc}).   In the third run of Advanced LIGO and Virgo in 2019,  more binary black hole mergers are detected (\cite{abbott2018prospects}).   In astrophysics,  how these binary compact objects form is still an open issue.   Usually,  there are two main channels: isolated binary evolution and dynamical formation.   These two channels admit the binary merging due to gravitational radiation within the cosmological age.   

The mechanism of formation of binaries does not directly encode into the GW signals.   However,  the imprint of the formation channels may be the eccentricity of the binary orbit. There are viable formation channels include isolated binary evolution (\cite{bethe1998evolution,belczynski2002comprehensive,belczynski2014formation,belczynski2016first,spera2015mass}) and dynamical encounters (\cite{zwart1999black,o2006binary,sadowski2008total,downing2010compact,downing2011compact}). Due to the GW radiation,  at the last stage of the merger,  the orbit will be definitely circularized (\cite{hinder2008circularization}). This is the reason why all events observed now have zero eccentricity. In the earlier stage,  for example for a binary black holes with the radiated GW frequency at 10 Hz ,  the eccentricity should be negligible if the merger comes from the isolated binary evolution (\cite{peters1964gravitational,hinder2008circularization}).   In the other case,  if the merger originates from dynamical formation,  the eccentricity could be in a wide range when radiated wave of binary at 10 Hz,  even can be close to unity (\cite{zevin2017constraining,zevin2019eccentric,zevin2019can,rodriguez2018triple,samsing2018eccentric,gondan2019measurement}).   These researches also predicted that about $5\%$ of dynamically-formed binaries to have $e \geq 0.1$ at 10 Hz (\cite{peters1963gravitational}).    

In the first and second runs of advanced LIGO and Virgo,  the starting GW frequency observed is more than 20 Hz (\cite{abbott2016characterization}).   However,  with the constant updating of LIGO,  and the future Einstein telescope (\cite{anselin1995local,aso2013interferometer,hu2017taiji,abbott2018prospects}),  we can observe the GWs at 10 Hz or even lower for coalescence of binary compacts.   So,  we expect that we can find the eccentric orbit at lower frequency band.   After enough events accumulated,  the distribution of two channels may be revealed.   

Usually,  the search of eccentric GW sources needs waveform templates.   Now in the LALsuite Liabray  (\cite{vallisneri2015ligo}),  there are EccentricFD (\cite{huerta2014accurate,tiwari2016proposed}) and TaylorF2e (\cite{moore20193pn}), together with ready-to-use eccentric model (\cite{tiwari2019ready}) which only cover the inspiraling part of binary merger.   For BBH mergers,  highly accurate models with the full inspiral- merger-ringdown,  along with support for both a large range of eccentricity there are models which satisfy some of these constraints (\cite{huerta2018eccentric,cao2017waveform,hinderer2017foundations,hinder2018eccentric,ireland2019eccentric}).   The SEOBNRE (\cite{cao2017waveform}) includes spin and has very good consistence with numerical relativity data.   

Before Advanced LIGO detected its first event GW150914, \cite{tiwari2016proposed}  proposed how to search the eccentric binary black holes.    \cite{abbott2019search}  announced the search result for eccentric binary black hole mergers with Advanced LIGO and Advanced Virgo during their first and second observing runs,  no candidate events were observed.   In the same year,   \cite{nitz2020search}   searched for eccentric binary neutron star mergers in the first and second observing runs of Advanced LIGO with matched filtering technology by using EccentricFD model,  and also no candidates were reported.   \cite{lower2018measuring,romero2019searching}  using SEOBNRE to search eccentricity in the first gravitational transient catalogue of LIGO and Virgo (\cite{abbott2019gwtc}),  they tried to perform Bayesian inference to measure the possible eccentricities while these events at 10 Hz.   They believe all the eccentricities should be zero even at the 10 Hz band.  

However,  due to the very high noise at the 10 Hz band (\cite{harry2010advanced,abbott2016characterization,martynov2016sensitivity}),  the above inference may not exclude the possibility of eccentricity at this low frequency stage,  as they said,  their analysis just yields no strong evidence for non-zero eccentricity in GWTC-1 (\cite{abbott2019tests}).   In the present paper,  we use the SEOBNRE which proposed by one of the authors to do a theoretical constrain of the eccentricities of GWTC-1 events (\cite{abbott2019search}).   We find that due to the fast circularization of the orbits by gravitational radiation (\cite{redmount1989gravitational,will1996gravitational,hinder2008circularization}),  though we observe zero eccentricity at $\gtrsim$ 20 Hz,  but a big range of eccentricity distribution at 10 Hz is still theoretically allowed.   

This paper are organized as follows.   In the next section,  we briefly introduce the SEOBNRE waveform model and demonstrate its performance on the modeling of eccentric waveforms.   In the third section,  by using SEOBNRE,  we generate eccentric waveforms at 10 Hz and do matched filtering with all BBH events announced in GWTC-1.   Finally,  discussion will be addressed in the last section.    

\section{The performance of SEOBNRE}
\subsection{The SEOBNRE model}

In order to describe the binary black holes, we use following parameters.  The masses of a binary black hole are $m_1$ and $m_2$ and we assume $m_1\geq m_2$, the total mass is $M=m_1+m_2$, the mass ratio is $q\equiv m_1/m_2$, in our paper, q is always bigger than 1.  The symmetric mass ratio is $\eta=m_1m_2/M^2$.  The spin of the two black holes are $\Vec{S_1}$ and $\Vec{S_2}$.   Using units $c = G = 1$ in this section,  we define the dimensionless spin parameters
\begin{align}
\chi_i=\Vec{S_i}/m_i^2
\end{align}
with $\chi_i \in [-1, 1]$. In our paper, we only consider the spin aligning with the orbital angular momentum of the binary, i.e., $z$ direction.  

\comment{In spin-aligned model, a single effective spin parameter is defined as, 
\begin{align}
\chi_{\mathrm{eff}}=\frac{m_{1} \chi_{1}+m_{2} \chi_{2}}{M}
\end{align}
}
The core idea of effective-one-body (EOB) theory treats a real two-body system as an equivalent one-body problem.   \cite{buonanno1999effective} first proposed the effective-one-body approach for solving the problem of relativistic binary.   The EOB formalism is more accurate than post-Newtonian approximation in Taylor expansion.  The EOB theory can give the complete process of the merger of compact binary,  including the inspiral, merge and ringdown.   After that,  \cite{buonanno2007pan} built an effective-one-body numerical-relativity (EOBNR) waveform model  which combines the effective-one-body theory and numerical relativity data.   The updated EOBNR model is SEOBNR which is extended to the spinning black holes (\cite{barausse2011extending,taracchini2012prototype}).   SEOBNR has been proven to be useful for quasi-circular orbit without procession (\cite{lovelace2016modeling}).   \cite{cao2017waveform} extended SEOBNR to SEOBNRE for elliptic binary black hole merger, and has been used in a lot of data analysis to find eccentric sources (\cite{cao2017waveform,abbott2019search,ramos2019first,liu2019validating}).   In the present work,  we also employ the SEOBNRE to calculate GW waveforms with orbital eccentricity and to analyze the LIGO-Virgo GW data.  

The EOB formalism includes three independent but interacting parts: (1) a description of the conservation part of the dynamics process of the compact binary(the Hamiltonian); (2) the radiation-reaction force; (3) the asymptotic gravitational waveform emitted by the binary.  

The following is the simple summary of the conservation part for SEOBNRE.   In the Newtonian two-body problem,  we can equivalently use a "test particle" with a reduced mass to orbit the "center mass" $M$ ($M$ is the total mass of the two bodies).   The EOB theory extends the Newtonian idea to general relativity,  that is,  to find an equivalent external space-time metric of a binary.   In the EOB approach,  we reduce the conservative dynamics of two-body problem to a geodesic motion of an effective test particle in an effectively deformed Kerr spacetime.   

The EOB Hamiltonian can be written as (\cite{barausse2011extending,taracchini2012prototype})
\begin{align}
H &= M \sqrt{1+2 \eta\left(\frac{H_{\rm eff}}{M \eta}-1\right)}\\
H_{\rm eff} &= H_{\rm N S}+H_{\rm S}+H_{\rm S C}
\end{align}
where the details of $H_{N S}, H_{S}$ and $H_{S C}$ can be found in \cite{cao2017waveform}.  

We can write the motion equation based on the Hamiltonian,  
\begin{align}
    \dot{\vec{r}}=\frac{\partial H}{\partial \overrightarrow{\tilde{p}}},  \quad\dot{\vec{p}}=-\frac{\partial H}{\partial \vec{r}}.  
\end{align}
Now we introduce the gravitational wave part of the model,  the SEOBNRE model provides expressions for the $2, 2$ spin-weighted spherical-harmonic modes of the GW signal.   The inspiral waveform are decomposed into a quasi-circular part and an eccentricity part.   First,  the quasi-circular part is
\begin{align}
    h_{\ell m}^{(C)}=h_{\ell m}^{(N, \epsilon)} \hat{S}_{\rm e f f}^{(\epsilon)} T_{\ell m} e^{i \delta_{\ell m}}\left(\rho_{\ell m}\right)^{\ell} N_{\ell m} {\rm with}
    h_{\ell m}^{(N, \epsilon)}=\frac{M \eta}{R} n_{\ell m}^{(\epsilon)} c_{\ell+\epsilon} V_{\Phi}^{\ell} Y^{\ell-\epsilon, -m}\left(\frac{\pi}{2}, \Phi\right), 
\end{align}\\
Where $R$ is the distance between detector and source, $\Phi$ is the orbital phase, $Y^{\ell m}(\Theta, \Phi)$ are the scalar spherical harmonics.   Second,  the $(2,2)$ mode containing eccentric part is
\begin{align}
      h_{22} &=2 \eta\left[\Theta_{i j}\left(Q^{i j}+P_{0} Q^{i j}+P^{\frac{3}{2}} Q_{\text {tail }}^{i j}\right)\right. +P_{n} \Theta_{i j}\left(P_{n}^{\frac{1}{2}} Q^{i j}+P_{n}^{\frac{3}{2}} Q^{i j}\right)+P_{v} \Theta_{i j}\left(P_{v}^{\frac{1}{2}} Q^{i j}+P_{v}^{\frac{3}{2}} Q^{i j}\right)\\&+P_{n n} \Theta_{i j} P_{n n} Q^{i j}+P_{n v} \Theta_{i j} P_{n v} Q^{i j}+P_{v v} \Theta_{i j} P_{v v} Q^{i j}+P_{n n v} \Theta_{i j} P_{n v}^{\frac{3}{2}} Q^{i j}+P_{n v v} \Theta_{i j} P_{v v v}^{3} Q^{i j}\left. +P_{v v v} \Theta_{i j} P_{v v v}^{\frac{3}{2}} Q^{i j}\right]\,.\notag
\end{align}
All the coefficients listed in the above equations can be found in (\cite{cao2017waveform}). 

\subsection{Validation of elliptic waveforms with numerical relativity data}
In this subsection,  we compare the SEOBNRE waveforms with NR data to validate this theoretical waveform model,  because description of a binary black hole by SEOBNRE depends on several approximations while the NR waveform is the direct solution to the Einstein equation. It is a standard procedure that using the numerical-relativity(NR) waveform to validate an approximated waveform (\cite{baumgarte2010numerical}).   The NR data we used are downloaded from \href{https://www.black-holes.org/code/SpEC.html}{https://www.black-holes.org/code/SpEC.html} (\cite{mroue2013catalog,blackman2015fast,boyle2019sxs}).  The quantitatively comparison  between the approximate waveforms ($h_1$) and the numerical-relativity waveforms ($h_2$) is using the standard inner product weighted by $S_n (f)$(the power spectral density of the detector noise,  here we use the LIGO's sensitivity curve).  

The SEOBNRE can generate the complete waveform include inspiral,  merge and ringdown.   The waveform of binary coalescence has an amplitude peak,  usually this moment is labeled as $t = 0$.   We align the NR and SEOBNRE waveforms at the amplitude peak to do the comparison.   The inner product is defined as (\cite{cutler1994gravitational}), 
\begin{align}
    \left\langle h_{1},  h_{2}\right\rangle= 4 \operatorname{Re} \int_{f_{\min }}^{f_{\max }} \frac{\tilde{h}_{1}(f) \tilde{h}_{2}^{*}(f)}{S_{n}(f)} d f
\end{align}\\
The normalized match optimized over a relative time shift and the initial orbital phase can be written as follow, 
\begin{align}
    M\left(h_{1},  h_{2}\right)=\max \left[\frac{\left\langle h_{1} | h_{2}\right\rangle}{\sqrt{\left\langle h_{1} | h_{1}\right\rangle\left\langle h_{2} | h_{2}\right\rangle}}\right]
\end{align}

For a given NR waveform, we set the parameters like as total mass$M$,  mass ratio $q$ and the individual spin $\chi_{1, 2}$.  We evaluate the SEOBNRE waveform between a frequency range of 20 and 2000Hz.  When we calculate the theoretical waveforms, we use the same $M, ~q, ~\chi$ of the NR waveform, but do not use the NR eccentricity to set the eccentricity in SEOBNRE.  Because the eccentricity changes with GW's (or orbital) frequency and the orbit is not closed ellipse,  the eccentricity in NR waveform may be not rigid.   

\begin{figure}
\begin{center}
\includegraphics[height=2.6in]{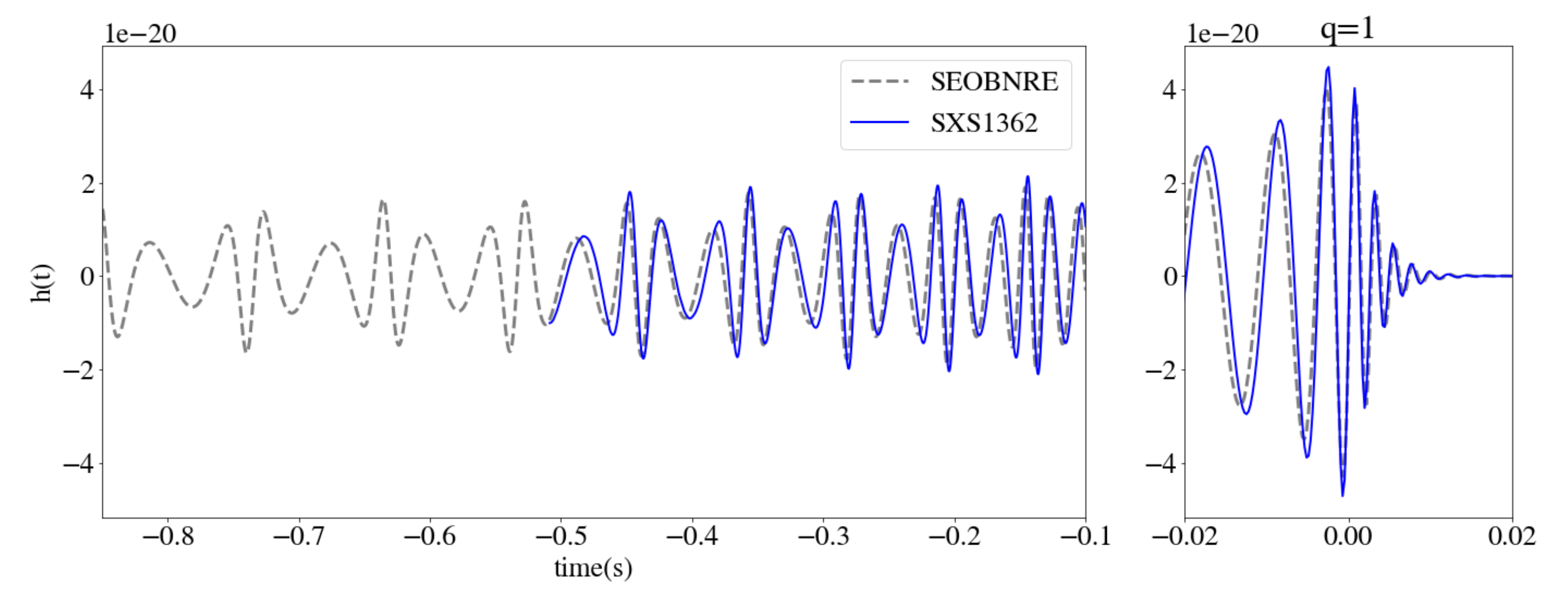}
\caption{NR (BBH:SXS1362) and SEOBNRE waveforms for equal-mass spinless BH binary with $m_1 = m_2 =20 m_\odot$.   The initial eccentricity of SXS1362 is $e < 1.7$ ,  and the one of SEOBNRE waveform is 0.401 at 20Hz.   The match between the two waveforms is 0.9905.}  \label{waveform1362}
\end{center}
\end{figure}

In Fig. \ref{waveform1362},  we plot both the SEOBNRE and NR waveforms for an equal-mass spinless BBH case (BBH:1362 in SXS data, $m_1 = m_2 = 20 m_{\odot}$) without spin. By setting the initial eccentricity of SEOBNRE equals 0.401 at 20 Hz, we get the best coincidence of two waveforms, and the match of them is 0.9905.  

\begin{figure}
\begin{center}
\includegraphics[height=3.0in]{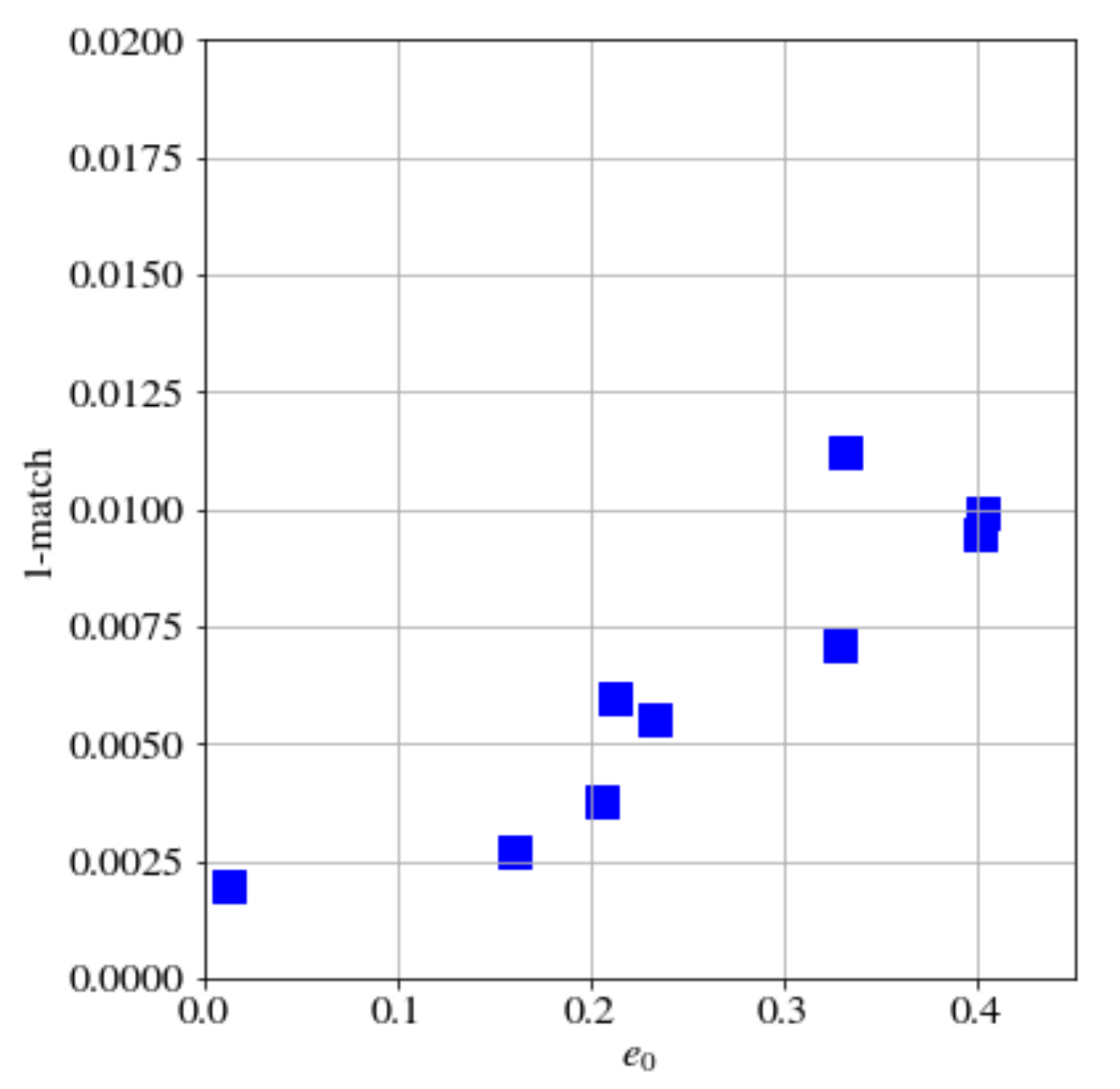}
\caption{The mismatch between the NR waveforms and SEOBNRE waveforms for equal-mass and nonspinning binaries, $e_0$ represents the initial eccentricity of the SEOBNRE waveform when the GW frequency is 20Hz as assuming $m_1 = m_2 = 20 m_\odot$.}  \label{matchq1}
\end{center}
\end{figure}

Fig. \ref{matchq1} shows the match results of a few NR waveforms with SEOBNRE ones for equal mass and nonspinning binaries.  The match results keep good with the increasing of the eccentricity.  Most of matches are better than 99\% even for the large eccentricity.  However, the mismatch still increases a little when the eccentricity grows bigger. This is clearly showed in the figure.  

Now we investigate the performance of SEOBNRE for general BBHs with spins or varied mass-ratios.  For this target, we use 14 NR BBH waveforms(SXS:320-324, SXS:1364-1373). Among them, SXS:320-324 are spin-aligned BBHs, with $\chi_1=0.44$, $\chi_2=-0.33$, and the mass ratio $q=1.22$. The eccentricities of these BBHS are in a range [0, 0.3]. SXS:1364-1373 are nonspinning binaries but the mass ratio is 2:1 or 3:1. In Table \ref{FFlist}, we list the match results of NR data with SEOBNRE waveforms in the third column. One can see that except for SXS 1361, all matches of the other cases are larger than 0.98. Together with the results in Fig. \ref{matchq1}, we have enough confidence for the SEOBNRE template in the cases of spinning and unequal mass-ratio BHBs with eccentricities. In addition, we also compare the eccentricities in the NR data and the ones in the SEOBNRE model.

We also compare the eccentricities defined in NR waveforms and SEOBNRE ones. Most of the eccentricities coincide each other in an acceptable errors, see Fig. \ref{1-match} for details. We notice that the initial eccentricities of SEOBNRE waveforms ar usually larger than the ones in NR data when the mass-ratio is 1.22 and 2.  However, when mass ratio equals one, the eccentricities of SEOBNRE waveforms are a little smaller than the NR ones. For some NR cases without physical eccentricities (such as 324, 1362 and etc.), SEOBNRE can offer reference values. 

\begin{figure}
\begin{center}
\includegraphics[height=2.6in]{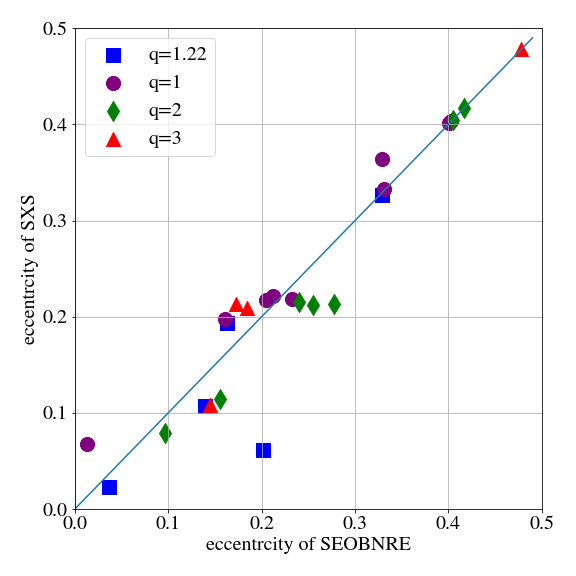}
\caption{comparison the eccentricity of SEOBNRE at 20Hz and the eccentricity given by the NR waveform, using different color to distinct different mass ratio, the mass ratio is indicated on the label, the eccentricity of SEOBNRE is all bigger when the mass ratio is 2 and 3.  when mass ratio q=1, the eccentricity of SEOBNRE is a little smaller than the NR's eccentricity}  \label{1-match}
\end{center}
\end{figure}

 There are several gravitational wave templates containing orbital eccentricity which have been implemented in LIGO Libraries and LALSuite, such as eccentricity FD and TaylorF2 ecc, which are non-spinning frequency domain waveform templates for eliptic binaries \cite{tanay2016frequency, huerta2014accurate}.  They both only describe the inspiraling part of the gravitational waves without the merger and ringdown waveforms. Considering the eccentricity FD and TalorF2 ecc models are inspiral-only templates, so there is a sharp cutoff in the end of the waveform.  For matching with NR data, we need cut the NR waveforms to keep only the inspiral part.

The match results of these two templates are also shown in Table I and Fig. \ref{1-match}.  The Table lists all the information about the BBH number of the SXS, the eccentricity of SXS, the match between SEOBNRE waveforms and SXS ones, the eccentricity of SEOBNRE at 20Hz, the match between eccentricity FD and SXS data, the match between TaylorF2 ecc and SXS data.   Fig. \ref{1-match} shows all the matches of three templates with NR data. From these results, we can see that the SEOBNRE performs much better than the other two templates when the BBHs having spins, asymmetric mass-ratio and nonzero eccentricities.  

\begin{table}
\centering
  \caption{Match results between SXS numerical relativity data with the SEOBNRE waveforms, eccentricity FD and TalorF2 ecc templates.} \label{FFlist}
  \begin{tabular}{c|c|c|c|c|c}
  \hline\hline
   SXS number & SXS ecc.   & SEOBNRE match & SEOBNRE ecc.   & eccFD match & TaylorF2 match\\
  \hline
  $320$& $0.0227$ &$0.9985$ &$0.037$ &$0.9839$ &$0.9820$\\
  $321$& $0.0611$ &$0.999$  &$0.202$ &$0.9761$ &$0.9841$\\
  $322$& $0.1070$ &$0.9983$ &$0.139$ &$0.9914$ &$0.9803$\\
  $323$& $0.1936$ &$0.9917$ &$0.163$ &$0.9504$ &$0.9558$\\
  $324$& $NaN$	  &$0.9986$ &$0.329$ &$0.9412$ &$0.9421$\\
  $1355$&$0.0678$ &$0.998$  &$0.013$ &$0.9587$ &$0.9663$\\
  $1356$&$0.1974$ &$0.9873$ &$0.161$ &$0.9598$ &$0.9600$\\
  $1357$&$0.2211$ &$0.9940$ &$0.212$ &$0.9512$ &$0.9517$\\
  $1358$&$0.2186$ &$0.9861$ &$0.233$ &$0.9379$ &$0.9384$\\
  $1359$&$0.2177$ &$0.9962$ &$0.205$ &$0.9650$ &$0.9666$\\
  $1360$&$0.3635$ &$0.9829$ &$0.241$ &$0.9855$ &$0.9863$\\
  $1361$&$0.3326$ &$0.9483$ &$0.212$ &$0.8882$ &$0.8884$\\
  $1362$&$<1.7e$  &$0.9905$ &$0.309$ &$0.9797$ &$0.9796$\\
  $1363$&$<1.8e$  &$0.9901$ &$0.314$ &$0.9772$ &$0.9777$\\
  $1364$&$0.0793$ &$0.9981$ &$0.097$ &$0.9680$ &$0.9849$\\
  $1365$&$0.1141$ &$0.9983$ &$0.155$ &$0.9738$ &$0.9751$\\
  $1366$&$0.2154$ &$0.9942$ &$0.24$  &$0.9230$ &$0.9373$\\
  $1367$&$0.2132$ &$0.9985$ &$0.277$ &$0.9844$ &$0.9787$\\
  $1368$&$0.2121$ &$0.9966$ &$0.255$ &$0.9510$ &$0.9612$\\
  $1369$&$<1.8$   &$0.9915$ &$0.405$ &$0.9867$ &$0.9848$\\
  $1370$&$<1.7$   &$0.9863$ &$0.417$ &$0.9755$ &$0.9755$\\
  $1371$&$0.1086$ &$0.9908$ &$0.145$ &$0.9637$ &$0.9638$\\
  $1372$&$0.2137$ &$0.9836$ &$0.173$ &$0.9450$ &$0.9451$\\
  $1373$&$0.2086$ &$0.9932$ &$0.184$ &$0.9299$ &$0.9329$\\
  \hline\hline
  \end{tabular}
\end{table}

\begin{figure}
\begin{center}
\includegraphics[height=2.6in]{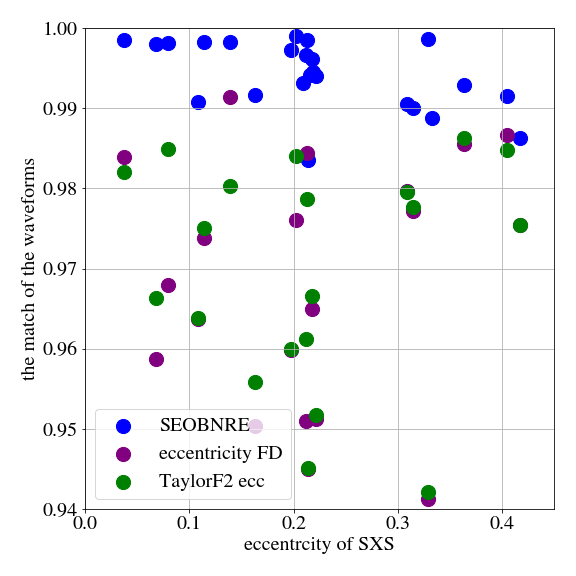}
\caption{Match results between SXS numerical relativity data with the SEOBNRE waveforms, eccentricity FD and TalorF2 ecc templates.} \label{1-match}
\end{center}
\end{figure}

 Because the eccentricity FD and TaylorF2 ecc waveforms are shorter than the complete waveforms, then the signal-to-noises (SNRs) will be smaller than the SEOBNRE model. We calculate their optimal SNR and compare to the SEOBNRE one as follow:
\begin{align}
     {\rm SNR}= 4 \operatorname{Re} \int_{f_{\min }}^{f_{\max }} \frac{\tilde{h}_{1}(f) \tilde{h}_{1}^{*}(f)}{S_{n}(f)} d f
\end{align}
Considering SEOBNRE is a time-domain model, we do the FFT before we calculate the SNR, transferring the time-domain waveform to a frequency-domain one. The SNRs of four BHBs are shown in Table \ref{snrlist}, we can see the inspiral-merge-ringdown waveforms calculated by SEOBNRE have higher SNRs than the inspiral only waveforms (eccentricity FD and TalorF2 ecc). This suggests that we should use SEOBNRE to find the potential eccentric GWs in LIGO-Virgo data, if the efficiency of SEOBNRE have been improved. For calculation, we take the total mass($M$)as 40 solar mass, and the distance from the detector to the GW source as 100 Mpc. The source location ($\theta$, $\phi$)=0, the inclination($\iota$) and the azimuthal angle($\varphi$) are both setted as 0.  

\begin{table}
\centering
  \caption{The optimal matched filtering SNRs by using different waveforms, with NR data as ``real" signals.} \label{snrlist}
  \begin{tabular}{c|c|c|c}
  \hline\hline
   mass ratio  &SEOBNRE optimal SNR&eccFD optimal SNR &aylor F2 optimal SNR\\
  \hline
  1& $35.  6$  &$24.  5$  & 24.  5\\
  1.  22& $30.  0 $ & $23.  8$ &23.  8 \\
  2& $37.  0 $ & $21.  6$ &23.  42\\
  3 & $29.  4$ & $21.  4$ &21.  4\\
  \hline\hline
  \end{tabular}
\end{table}

\section{Estimating eccentricities at earlier stage of LIGO events} 
The advanced LIGO detectors began first run O1 on 2015 September 12.  Not long after that, the first gravitational signal GW150914 was detected.  The Advanced LIGO and Virgo has announced 12 gravitational wave events,  10 of them are coalescence of binary black holes.  All the gravitational wave sources have no eccentricity. This is because that maybe all the events come from isolated evolution, and the eccentricity is ignored before merger. However, maybe some of them have wild eccentricities at the earlier stage, and loss eccentricities due to the rapid circularizing at the final inspiraling stage when enter the LIGO's sensitive band. The circularization and the merger time scale can be estimated by Peters through post-Newtonian approximation in (\cite{peters1964gravitational})
\begin{align}
    \frac{d e}{d t}&=-\frac{304}{15} \frac{M^{3} \eta}{a^{4}\left(1-e^{2}\right)^{5 / 2}} e\left(1+\frac{121}{304} e^{2}\right)\\
    T &=\frac{768}{425} \frac{5 a_{0}^{4}}{256 M^{3} \eta}\left(1-e_{0}^{2}\right)^{7/2}
\end{align}

As we mentioned, a few of models predict that the eccentricity could be in a wide range when the radiated wave of binary at 10 Hz if the BHBs originates from dynamical formation. From the above equations, the eccentricity will be reduced to zero when GWs enter the detectable frequency ($>$ 20 Hz) of the first LIGO run. 

For investigating this possibility, we employ the SEOBNRE model and the binary black holes events released in the LIGO-Virgo catalog GWTC-1. Firstly, using the SEOBNRE, elliptic waveforms starting from 10 Hz are generated with the same parameters of the public BHB events.  Secondly, consider the sensitivity in the first run of LIGO, we only use the LIGO data of each event starting from 20 Hz.  Therefore, the SEOBNRE waveforms are used for data analysis also from 20 Hz, though the waveforms are calculated from 10 Hz with initial eccentricity. Finally, we match the theoretical waveforms and observed signals, calculate the matched-filtering SNR. 

\begin{figure}
\begin{center}
\includegraphics[height=1.6in]{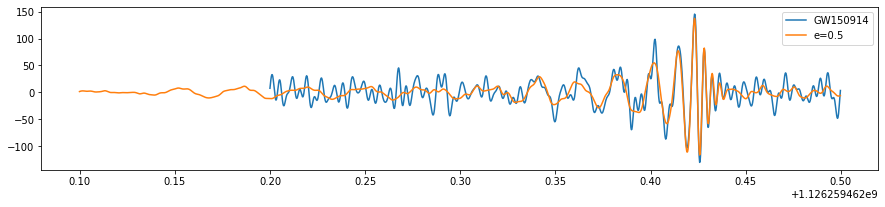}
\includegraphics[height=1.67in]{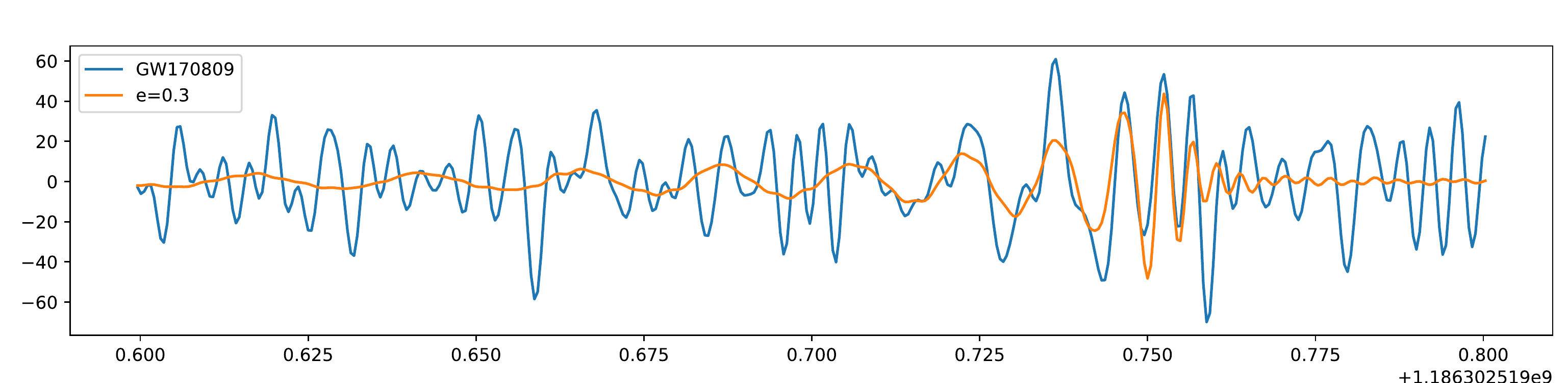}
\caption{Top panel: The SEOBNRE waveforms with $e_0 = 0.5$ at 10 Hz and GW150914 signals; Bottom panel: The SEOBNRE waveforms with $e_0 = 0.3$ at 10 Hz and GW170809 signals.}  \label{waveforms}
\end{center}
\end{figure}

In Fig. \ref{waveforms}, as examples, we demonstrate two SEOBNRE waveforms (noise added) together with the LIGO's signals GW150914 and GW 170809. With initial eccentricities of 0.5 and 0.3 at 10 Hz respectively, the SEOBNRE waveforms coincide with the two signals. This means that there is a possibility that this two BHBs allow eccentricities up to 0.5 and 0.3 at 10 Hz. 

After investigating all the ten BHB data, we find seven of them can not rule out possible eccentricities at 10 Hz. The matched-filtering SNRs with SEOBNRE templates of these seven events are larger than or equal the SNRs announced in LIGO-Virgo catalog.  All the data analysis are done by the software in LALsuite Liabray. In Table. \ref{ecclist}, we list the results of all these seven events, the second column list the maximal allowed eccentricities in these events with the same SNRs of LIGO released. 

\begin{table}
\centering
  \caption{The matched-filtering SNRs by using SEOBNRE templates of seven detected BBH events and possible maximum eccentricities at 10 Hz.} \label{ecclist}
  \begin{tabular}{c|c|c}
  \hline\hline
   Events  & possible $e_{\rm max}$ at 10 Hz & SNR  \\
  \hline
  GW150914& $0.5$  &$26.42$  \\
  GW170814& $0.08 $ & $17.6$ \\
  GW170104 &$0.08$ & $13.3$ \\
  GW170809 &$0.32$ & $12.69$\\
  GW170823 &$0.26$ & $11.38$\\
  GW170729 &$0.26$ & $10.42$\\
  GW170818 &$0.35$ & $10.47$ \\
  \hline\hline
  \end{tabular}
\end{table}

\begin{figure}
\begin{center}
\includegraphics[height=1.6in]{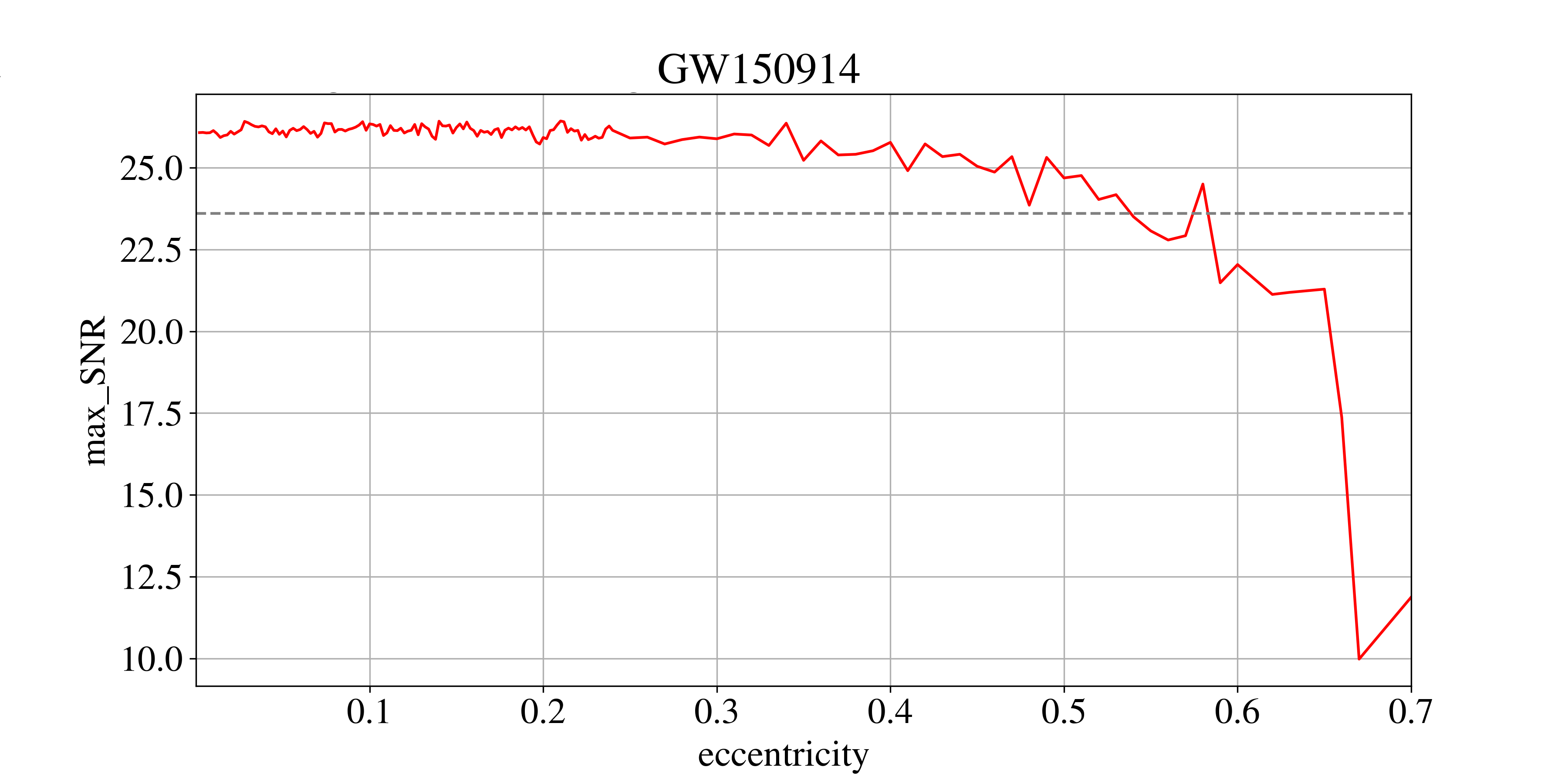}
\includegraphics[height=1.6in]{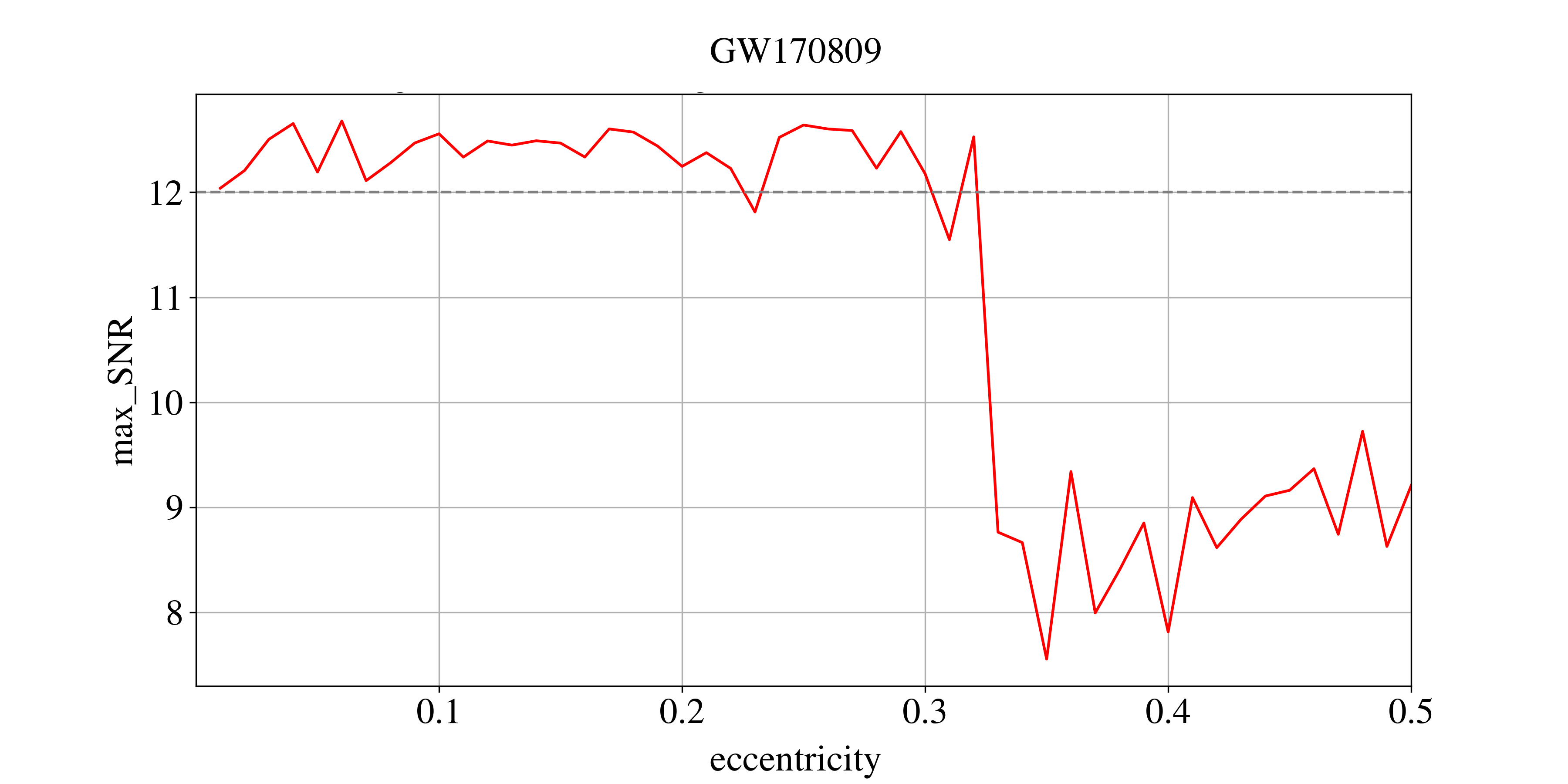}
\includegraphics[height=1.6in]{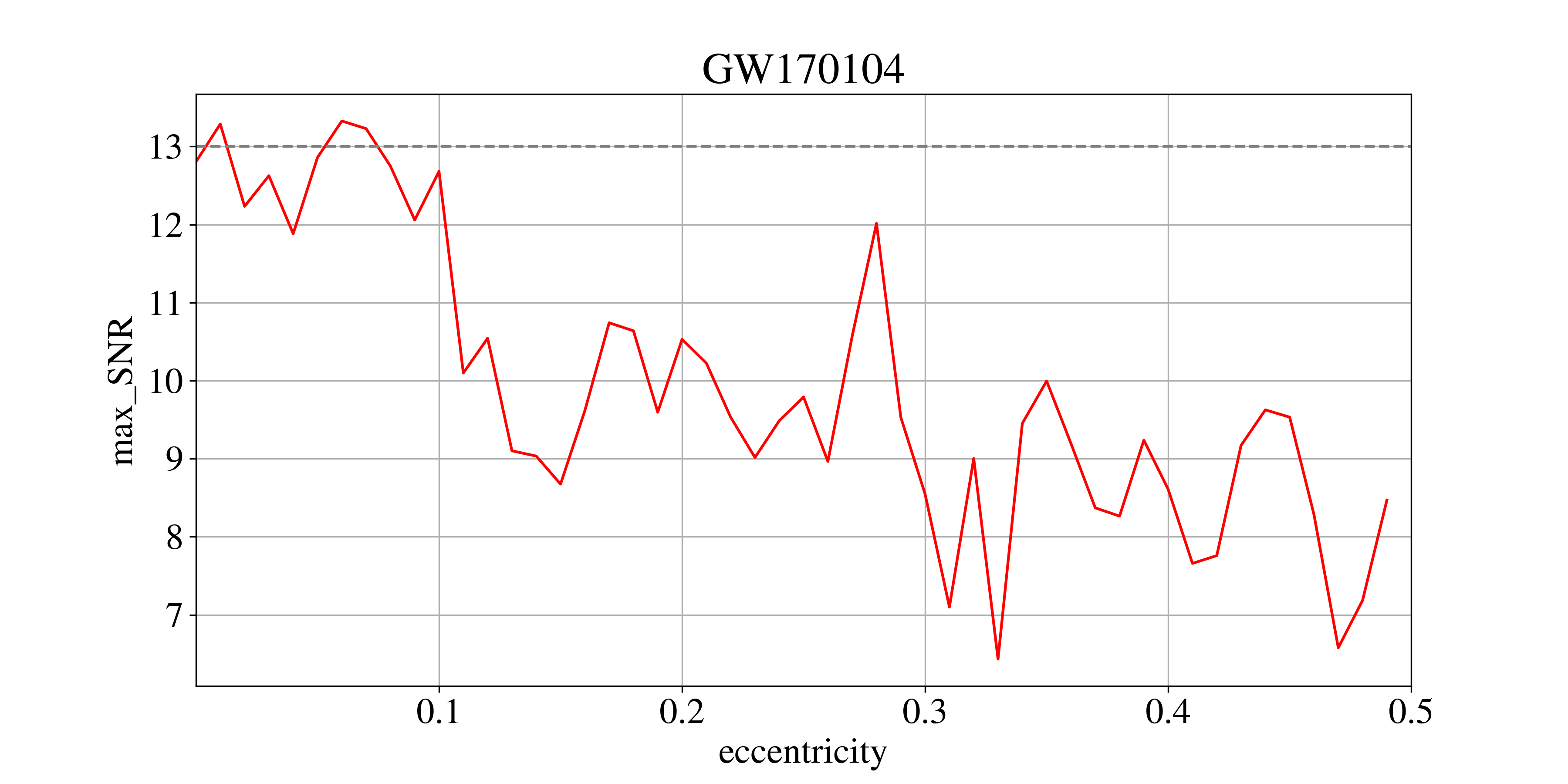}
\includegraphics[height=1.6in]{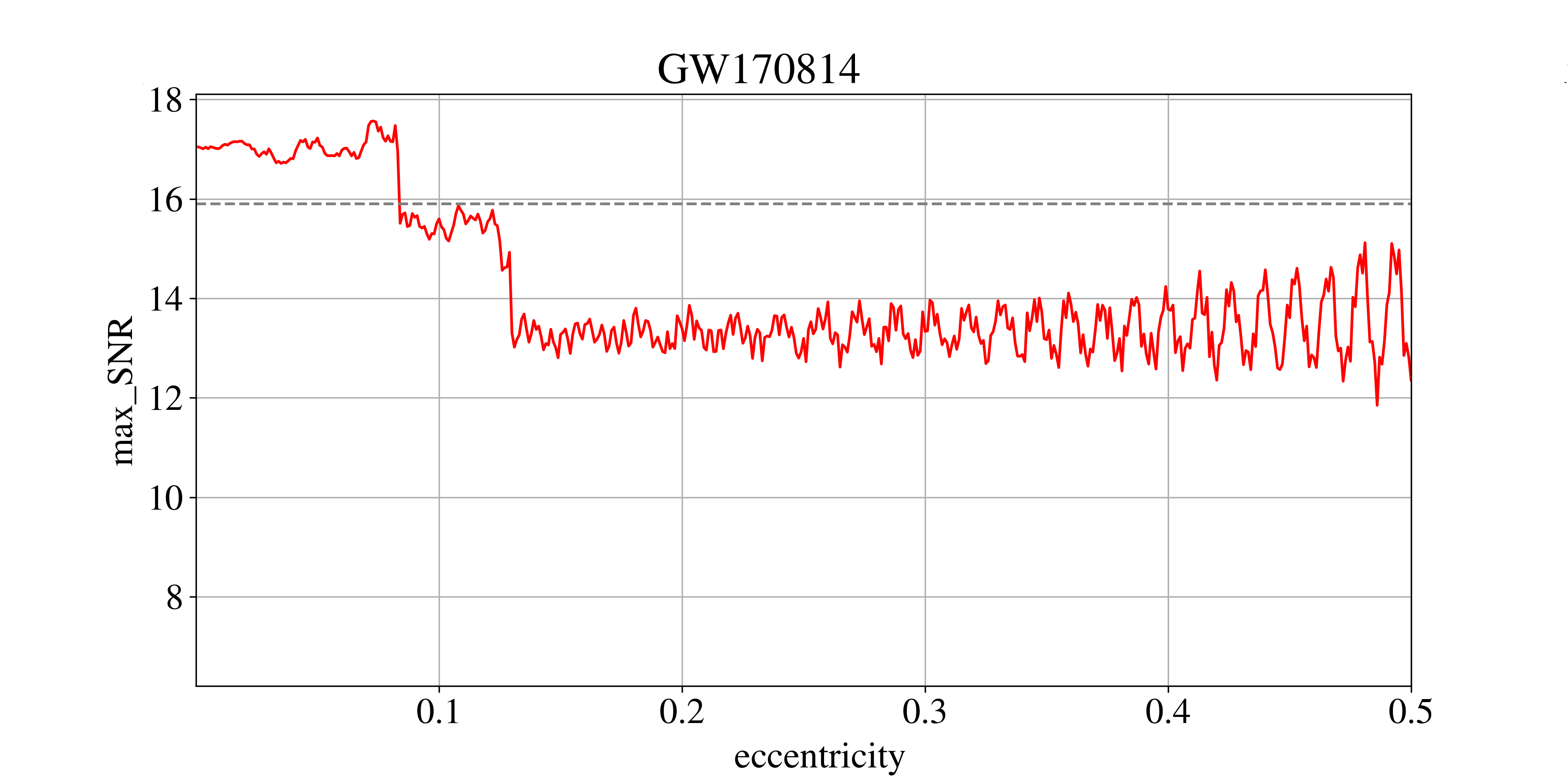}
\caption{Variation of the matched-filtering SNRs with initial eccentricities for four BHB events. The horizontal gray-dashed line is the LIGO official SNRs.}  \label{eccSNRs}
\end{center}
\end{figure}

In Fig. \ref{eccSNRs}, the variation of SNRs of four events (GW150914, GW 170104, GW170809 and GW170814) with initial eccentricities at 10 Hz is plotted. One can see that the SNRs drop suddenly if the initial eccentricities in the SEOBNRE waveforms exceed some critical values. We then assume that these events allow the possibility of orbital eccentricities at the 10 Hz stage. Of course, it is possible that all these events are still circular orbits at this earlier stage. 

From the above analysis, the possible range of initial eccentricities at 10 Hz of these seven BHB events is $[0,~e_{\rm max}]$. Our results only can declare that the current analysis can not rule out the possibility of ecliptic orbits at 10 Hz band. The initial frequency of the SEOBNRE waveforms with eccentricity is 10Hz, so all the initial eccentricity we talk about in this section is the eccentricity at 10Hz. 

\section{conclusions}
The formation mechanism of binary black hole mergers is an attractive topic in astrophysics, and eccentricity of binary is a useful clue to distinguish the origin of BHBs. Up to now, all LIGO-Virgo BHB events are announced as circular binaries. However, due to the poor sensitivity at relative low frequency band in the first run of LIGO, this declaration may be only completely correct for these GW events when they enter $\gtrsim 20$ Hz band, i.e., the final inspiral stage. 

In the present paper, by employing SEOBNRE, a full waveform templates including inspiral, merger and ringdown phases, we try to investigate the possibility that these BHB events may originate from ecliptic binaries when the GWs they radiated are at 10 Hz band. The choice of 10 Hz is based on two reasons. One is that the Advanved LIGO is improving its sensitivity after 10 Hz band, and maybe the data from this frequency will be available in the future. The other one is that some theoretical predictions show a few friction of  dynamically-formed binaries having nonzero eccentricity at 10 Hz.

By comparing with numerical relativity data, we validate that SEOBNRE is a reliable waveform model for eliptic binaries, and have better performance than the frequency-domain templates in LALsuite Library. We then use SEOBNRE to generate theoretical gravitational waveforms from 10 Hz with eccentricity from 0 to 0.7. With the matched-filtering technology, we analysis the LIGO data of ten BHB events in the first catalog GWTC-1, and find that elliptic waveforms with initial eccentricities at 10 Hz can match very good with the observed signals after 20 Hz for seven BHB events.  

Therefore, we conjecture that all these seven events still allow to exist eccentricities at the earlier stage (10 Hz band). However, it must be emphasised that we DO NOT measure the eccentricities at 10 Hz, and of course we DO NOT announce that these events are elliptic at this frequency band. Our results just show that nonzero eccentricities at 10 Hz of these events are possible. Due to the orbital circularization by gravitational radiation, the eccentricity reduces to zero after the GW frequency goes into 20 Hz.


In the near future, as long as the update of Advanced LIGO and Virgo, GW data from 10 Hz will be available. One then can really measure the eccentricities of binary mergers at this frequency band. The detection of an eccentric binary can not only prove the binary can form dynamically but also distinguish the different dynamical formation (dynamical encounter or Kozai-Lidov oscillations in triple systems). 




\section*{Acknowledgements}

This work is support by NSFC No. 11273045, No. 11773059 and by Key Research Program of Frontier Sciences, CAS, No. QYZDB-SSW-SYS016. We also appreciate the useful discussions with Prof. Zhoujian Cao and Xiao-Lin Liu. This work was also supported by MEXT, JSPS Leading-edge Research Infrastructure Program, JSPS Grant-in- Aid for Specially Promoted Research 26000005, JSPS Grant-in-Aid for Scientific Research on Innovative Areas 2905: JP17H06358, JP17H06361 and JP17H06364, JSPS Core-to-Core Program A. Advanced Research Networks, JSPS Grant-in-Aid for Scientific Research (S) 17H06133, the joint research program of the Institute for Cosmic Ray Research, University of Tokyo.





\bibliographystyle{mnras}








\bsp	
\label{lastpage}
\end{document}